# Energy dissipation in van der Waals two-dimensional devices

**Zhun-Yong Ong[1] and Myung-Ho Bae[2,3*]**

[1]Institute of High Performance Computing, #16-16, 1 Fusionopolis Way, Agency for Science, Technology and Research, 138632, Singapore.

[2]Korea Research Institute of Standards and Science, Daejeon 34113, Republic of Korea

[3]Department of Nano Science, University of Science and Technology, Daejeon, 34113, Republic of Korea

[*]e-mail address: mhbae@kriss.re.kr

**Abstract**

Understanding the physics underlying energy dissipation is necessary for the effective thermal management of devices based on two-dimensional (2D) materials and requires insights into the interplay between heat generation and diffusion in such materials. We review the microscopic mechanisms that govern Joule heating and energy dissipation processes in 2D materials such as graphene, black phosphorus and semiconducting transition metal dichalcogenides. We discuss the processes through which non-equilibrium charge carriers, created either transiently through photoexcitation or at steady state by a large electric field, undergo energy relaxation with the lattice and the substrate We also discuss how these energy dissipation processes are affected by the device configuration (heterostructure, substrate material including hexagonal boron nitride, etc) as the use of different substrates, encapsulation, disorder, etc can introduce or remove scattering processes that change the energy relaxation pathways. Finally, we discuss how the unique carrier scattering dynamics in graphene-based vdW heterostructures can be exploited for optoelectronic applications in light emission and photodetection.

## 1. Introduction

One of the central themes in the development of the modern semiconductor device technology is the progressive miniaturization of field-effect transistors (FETs) and other silicon-based device components. With device feature sizes reaching ~10 nm [1], energy efficiency problems associated with power consumption and energy dissipation have become significant. Some of the factors directly affecting power consumption are the leakage currents caused by the short channel effect and electron tunneling through thin gate insulators [2]. To address this issue, alternatives to conventional semiconductor materials, in the form of two-dimensional (2D) materials such as semi-metallic graphene and semiconducting transition-metal dichalcogenides (TMDCs) [3], have been explored. Since these two-dimensional materials are only a few times thicker than an atomic layer, they can potentially solve the short-channel effects that conventional silicon cannot overcome for the same given channel thickness [4], and their inherent properties do not change even when they are bonded with other materials by van der Waals (vdW) interaction. Due to these advantages, the development of vdW low-dimensional devices in which 2D materials are integrated with other 2D materials or other conventional semiconductor materials has accelerated in the last decade. However, one of the issues that must be taken into consideration in the development of these materials for device technology is energy dissipation arising from charge carrier scattering processes [5] which transfer energy from the driving electric field to the crystal lattice, generating waste Joule heat that can increase the local temperature of the 2D materials and lead to suboptimal device performance In addition to energy loss within the material itself during device operation, the energy dissipation is also greatly influenced by the interaction with the environment due to the thickness of the atomic layer of the material itself. Hence, a deeper understanding the microscopic mechanisms that govern Joule heating and energy dissipation processes in these 2D materials can lead to superior thermally aware designs and effective thermal management of semiconductor devices. Achieving this requires insights into the interplay between heat generation from electron-phonon interactions and heat conduction through diffusion of the thermally generated phonons.

In this review, we provide a broad overview of the physics underlying energy dissipation in low-dimensional device structures. In Section 2, we first discuss thermal properties of 2D materials such thermal conductivity and thermal boundary resistance. In Section 3, we mainly examine the energy dissipation in steady states of operating graphene devices including TMDCs, Black phosphorous (BP) and vdW heterostructures. Close attention is paid to the experimental characterization of devices under high power densities. Finally, in Section 4, we deal with the energy relaxation of photoexcited carriers in graphene and vdW heterostructures in dynamic regimes.

## 2. Thermal properties of two-dimensional layered materials

Although the van der Waals (vdW) heterostructure should be properly regarded as a composite, its thermal properties can often be understood or at least interpreted in terms of those of its constituent materials. In this section, we give an overview of the thermal transport properties of the individual two-dimensional (2D) layered materials which form the basic building blocks in van der Waals heterostructures, with particular attention paid to aspects which are relevant for energy dissipation within the heterostructure, such as the directional dependence of the thermal conductivity. Four 2D materials – graphene, hexagonal boron nitride (hBN), molybdenum disulphide ($MoS_2$) and black phosphorus (also known as phosphorene) are reviewed here as they are representative of the wider class of 2D materials that are used in vdW heterostructure research. We briefly discuss the similarities and differences in their thermal conductive properties and relate them to their crystal structure. Due to length constraints, no attempt is made to delve into the finer theoretical aspects of the thermal properties of individual 2D crystals. Rather, we focus our attention on the major issues in the thermal properties of 2D materials that are significant for understanding energy dissipation in individual 2D materials and heterostructures. For a more detailed and comprehensive exposition, we refer the reader to Gu and Yang's excellent review article [6] as well as other recent reviews [7,8]. Specific reviews of the thermal transport properties of individual 2D materials can also be found in Ref. [9] (graphene) and Ref. [10] (black phosphorus).

## 2.1. Thermal conductivity physics

Before we discuss the differences in thermal properties, it is necessary to briefly review some of the fundamental physical concepts underlying heat conduction in 2D materials. In crystalline semiconductors and insulators, a consequence of the low charge carrier density is that the electronic contribution to thermal conductivity is insignificant and thermal transport is primarily mediated by phonons, quantized lattice vibrational motion, of which there are three primary branches: one longitudinal and two transverse acoustic. In the general case where acoustic phonon group velocities are direction-dependent and the thermal conductivity is anisotropic, the thermal conductivity tensor, which characterizes heat conduction, can be expressed in the relaxation time approximation as the sum of the acoustic phonon contributions, i.e.,

$$\kappa_{\alpha\beta} = \sum_{l} c_l v_{l\alpha} v_{l\beta} \tau_l \tag{1.1}$$

where for phonon mode $l$, $c_l$, $v_{l\alpha}$ and $\tau_l$ are respectively its heat capacity, phonon group velocity in the α-direction and relaxation time, a measure of how long it takes for the nonequilibrium phonon distribution to relax back to its equilibrium distribution. The *direct* contribution by optical phonons to the thermal conductivity in Eq. (1.1) is typically ignored because of their very low group velocities and thermal population at room temperature, although they influence the relaxation times of the acoustic phonons through Umklapp anharmonic scattering.

The heat flux $q_\alpha$ is related linearly to the temperature gradient $\partial T / \partial x_\beta$ via the expression

$$q_\alpha = -\sum_{\beta} \kappa_{\alpha\beta} \frac{\partial T}{\partial x_\beta} \tag{1.2}$$

and it depends on the orientation of the temperature gradient with respect to the crystal axes, of which there are three independent components along the principal axes of the thermal conductivity tensor. In a layered crystal such as graphite, the thermal conductivity tensor is highly anisotropic along these axes, with the cross-plane (*c*-axis) component being much smaller than the other two because of the weak interlayer coupling and low phonon group velocities in the direction of the *c*-axis. Apart from the phonon group velocities which determine how fast the phonons propagate, the thermal conductivity in Eq. (1.1) also depends on the relaxation times which can be calculated from individual mode-dependent scattering rates, *i.e.*

$$\tau^{-1} = \tau_{anh}^{-1} + \tau_{def}^{-1} + \tau_{sur}^{-1} \tag{1.3}$$

where $\tau_{anh}^{-1}$, $\tau_{def}^{-1}$ and $\tau_{sur}^{-1}$ are the scattering rates associated with phonon anharmonicity, defect scattering and surface scattering, respectively. In general, variation in the thermal conductivity between nonmetallic crystalline solids can be explained by the differences in phonon group velocities and scattering rates. Furthermore, variation in thermal conductivity is frequently characterized by the phonon mean free path $\lambda$, which is defined as the product of the phonon group velocity and relaxation time and is used in the literature as a characteristic length scale for distinguishing ballistic and diffusive phonon transport.

**2.2 Phonon dispersion of 2D crystals**

In a freestanding 2D crystal such as graphene, the three acoustic phonon branches that dominate the thermal conductivity can be classified as either *in-plane* or *out-of-plane*, based on the direction of the atomic displacements in each mode. The in-plane acoustic phonons include the *longitudinal acoustic* (LA) and the *transverse acoustic* (TA) phonons, which are polarized within the plane, and they have a *linear* (ω ∝ q) dispersion near the gamma point of the Brillouin Zone. The LA phonons correspond to compressional waves in which the displacement of atoms is along the direction of wave propagation while the TA phonons correspond to in-plane shear waves in which the displacement of atoms is perpendicular to the wave vector. The remaining transverse acoustic phonon branch, commonly labeled as the flexural acoustic (ZA), is polarized in the out-of-plane direction, corresponding to the bending motion of the sheet, and has a *quadratic* ($\omega \propto q^2$) dispersion relationship when the sheet in unstrained although it has been claimed that tensile strain leads to its linearization [6,11]. However, it has been argued [12] that the coupling between the bending and stretching degrees of freedom in graphene can modify the quadratic dispersion and lead to a temperature-dependent $\omega \propto q^{3/2}$ dispersion relationship in the long-wavelength limit.

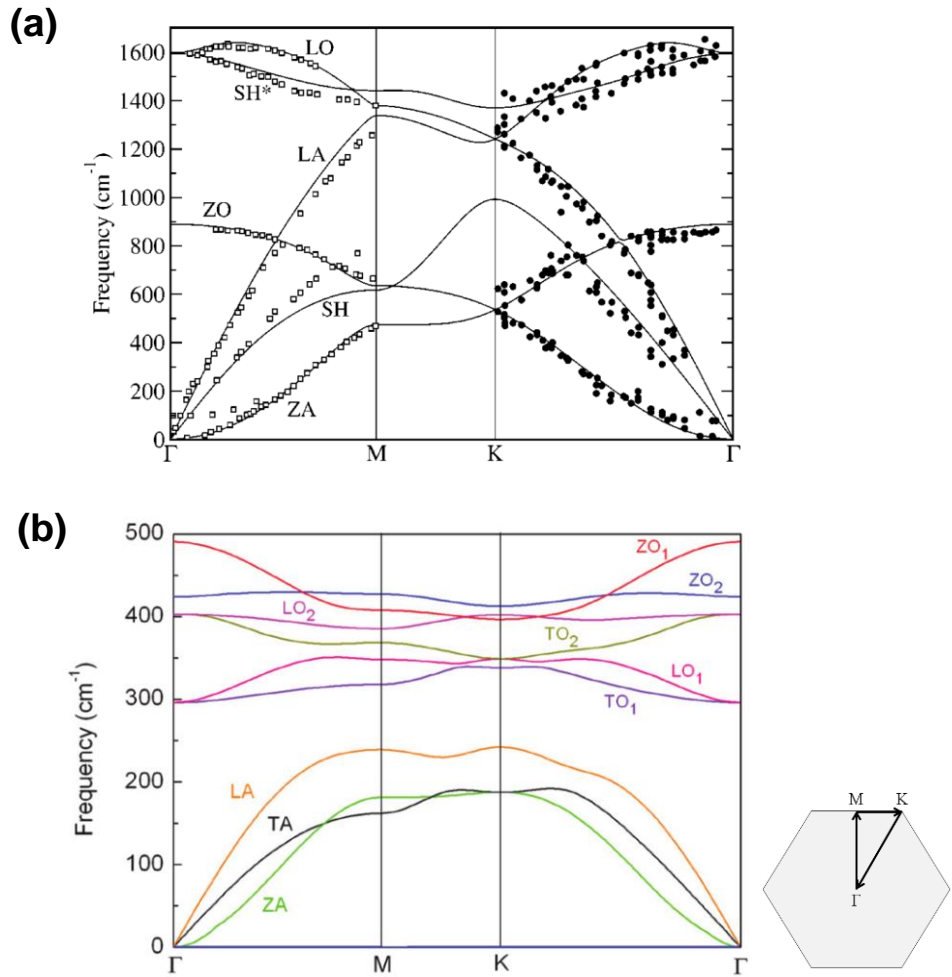

Figure 1. Plots of the phonon dispersion of (a) graphene and (b) MoS$_2$ calculated using density functional theory. Figure (a) is reproduced with permission from [13], © 2003 American Physical Society; (b) from [14], © 2014 American Physical Society.

Figure 1(a) shows the phonon dispersion relation for monolayer graphene taken from Ref. [13] while Fig. 1b shows the phonon dispersion relation for monolayer 2H-MoS$_2$ taken from Ref. [14]. Although graphene and MoS$_2$ have 6 and 9 phonon branches, respectively, as a result of their different number of atoms in the unit cell, their lowest three branches correspond to the LA, TA and ZA phonons and show the characteristic frequency convergence to zero at the $\Gamma$ point. In graphene and MoS$_2$, although the LA and TA phonons exhibit the expected linear relationship ($\omega_{LA} = c_{LA}q$ and $\omega_{TA} = c_{TA}q$) between frequency ($\omega$) and wave vector ($q$) where $c_{LA}$ and $c_{TA}$ are the LA and TA phonon group velocity, respectively, the acoustic phonon group velocities differ considerably between graphene and MoS$_2$. In the long-wavelength limit and the $\Gamma$-M direction, we have $c_{TA}$ = 693.5 m/s and $c_{LA}$ = 1108.8 m/s for MoS$_2$ and $c_{TA}$ = 3743 m/s and $c_{LA}$ = 5953 m/s for graphene. The substantially higher phonon group velocities in graphene are due to its much lighter C atoms relative to the heavier M and S atoms in MoS$_2$. Given Eq. (1.1), the much higher group velocities in graphene suggest that its thermal conductivity should be higher. Indeed, the differences in phonon group velocities between 2D materials can account for the disparities in thermal conductivity.

## 2.3 Thermal conductivity of 2D monolayers

### 2.3.1 Graphene

Graphene was first isolated and discovered in 2004 by Novoselov *et al* [15], and its electrical and thermal properties have been extensively characterized. In its monolayer form, pristine graphene consists of an atomically thin planar sheet of carbon atoms arranged in a honeycomb lattice with a two-atom unit cell. While its electronic properties have been the focus of much attention, a substantial amount of experimental and theoretical effort has also been directed towards investigating thermal transport in graphene. Indeed, because graphene is the archetype of the 2D material, we will discuss the physics underlying heat conduction in graphene in considerably more detail as the theoretical concepts introduced here will serve as the basis for understanding the microscopic mechanism of heat conduction and energy dissipation in the wider class of 2D crystals. In graphene, the thermal conductivity is expected to be very high given the relatively light C atoms and the strong C-C $sp^2$ bonds that lead to high LA and TA phonon group velocities. Experimental measurements of the room-temperature thermal conductivity yield a range of values between 600 and 5300 $Wm^{-1}K^{-1}$, with variability in thermal conductivity values due to differences in experimental interpretation and setup. The highest value reported (5300 $Wm^{-1}K^{-1}$) by Balandin *et al* [16] who measured the thermal conductivity of a suspended graphene sheet using the optothermal Raman technique [7], assuming an optical absorption of ∼13 percent. Cai *et al*, who measured a considerably lower optical absorption (∼3.3 percent), reported a smaller thermal conductivity of around 2500 $Wm^{-1}K^{-1}$ near 350 K although the lower reported value may be in part due to the use of CVD graphene [17]. On the other hand, the room-temperature thermal conductivity for substrate-supported graphene has been measured to be 600 $Wm^{-1}K^{-1}$ [18], an order of magnitude lower than the thermal conductivity in suspended graphene.

Although the large LA and TA phonon group velocities have been invoked to explain the high thermal conductivity in graphene [19], it has been argued by Lindsay and Broido that based on the large density of states for the out-of-plane ZA phonons and the plane-reflection symmetry-based selection rule restricting anharmonic phonon-phonon scattering of ZA phonons, the high lattice thermal conductivity of graphene is due primarily to the transport of ZA phonons which have significantly longer mean free paths [20]. At room temperature (300 K) in a graphene sheet of length $L = 10$ μm, the ZA phonon contribution is estimated to be between 70 and 80 percent of the overall lattice thermal conductivity, which is estimated to be around 3500 $WK^{-1}m^{-1}$, with the remainder coming from LA, TA and optical phonons [20]. The argument in favor of the dominant role of ZA phonons finds support in the markedly lower thermal conductivity of supported graphene relative to suspended graphene, which has been attributed to the damping of the ZA phonons by the substrate [18,21,22].

*2.3.2 Hexagonal boron nitride*

The lattice structure of single-layer hexagonal boron nitride (*h*-BN) is similar to that of graphene but with a two-atom primitive unit cell consisting of a B and N atom. Given its large band gap (∼5.8 eV) and its smooth planar surface, hexagonal boron nitride (hBN) is increasingly used as a dielectric substrate material as well as to encapsulate other 2D crystals in heterostructures used to probe their intrinsic electronic properties [23–25]. In spite of its widespread use, the thermal properties of *h*-BN have not been as well-studied as those of graphene. To the best of our knowledge at the time of writing, there has not been any experimental measurement of the thermal conductivity of monolayer h-BN although the thermal conductivity of multilayer h-BN has been characterized [26,27]. Using the microbridge method, Jo *et al* found $\kappa =$ 250 and 360 $Wm^{-1}K^{-1}$ for 5- and 11-layer *h*-BN at room temperature while Zhou *et al* obtained $\kappa =$ 280 Wm$^{-1}$K$^{-1}$ for 9-layer *h*-BN using the optothermal Raman technique, significantly lower than the thermal conductivity values obtained for single-layer graphene [17].

Given its structural similarity to graphene, we expect h-BN to exhibit the same reflection symmetry that is responsible for the reduction of anharmonic scattering of the ZA phonons [28]. Hence, theoretical calculations of the thermal conductivity of h-BN, estimated to be around 500 WK-m$^{-1}$ at room temperature in a 1 μm-long sheet, show that it is dominated by ZA phonons although the thermal conductivity is predicted to be significantly lower by an order of magnitude in *h*-BN than in graphene because of the stronger anharmonic phonon-phonon coupling in *h*-BN [28]. The markedly lower thermal conductivity in *h*-BN has also been reported by Sevik *et al* who used molecular dynamics simulations [29].

### *2.3.3 Molybdenum disulfide and other transition metal dichalcogenides*

Unlike graphene and h-BN, a single sheet of MoS$_2$ comprises three covalently bonded atomic layers, of which the top and bottom layers are S atoms while the middle layer consists of Mo atoms. There are two main polymorphs of MoS$_2$ (2H-MoS$_2$ and 1T-MoS$_2$) which differ from each other by the stacking order of the S-Mo-S atomic layers: the stacking order is *ABA* in 2H-MoS$_2$ and *ABC* in 1T-MoS$_2$. As a result of the difference in stacking order, 2H-MoS$_2$ and 1T-MoS$_2$ have substantially different electronic properties and thermodynamic stability [30]. The thermodynamically stable 2H-MoS$_2$ monolayer is a semiconductor with a band gap of 1.8 eV while the metastable 1T-MoS$_2$ monolayer is metallic. Given the thermodynamic instability and metallic nature of 1T-MoS2, there is much less experimental characterization of its thermal properties although there have been attempts to model the phonon part of its thermal conductivity. Thus, we focus our discussion on the thermal properties of 2H-MoS$_2$ and we drop the 2H prefix in all subsequent references to 2H-MoS$_2$.

Thermal transport in MoS$_2$ has been characterized mainly in few-layer samples. Sahoo *et al* obtained $\kappa =$ 52 Wm$^{-1}$K$^{-1}$ at room temperature in an 11-layer sample using the optothermal technique [31]. The room-temperature thermal conductivity of suspended monolayer MoS$_2$ was measured to be $\kappa =$ 34.5 Wm$^{-1}$K$^{-1}$ by R. Yan and coworkers [32]. The thermal conductivity of monolayer MoS$_2$ supported in a SiO$_2$ substrate was measured to be $\kappa =$ 62 Wm$^{-1}$K$^{-1}$ by A. Taube *et al* [33]. These thermal conductivity values for monolayer MoS$_2$ are comparable to theoretically predicted ones. Li and coworkers estimated $\kappa =$ 90 Wm$^{-1}$K$^{-1}$ for isotopically pure monolayer MoS$_2$ and $\kappa =$ 71 Wm$^{-1}$K$^{-1}$ for monolayer MoS$_2$ with naturally occurring M and S isotope percentages [34]. Using a similar approach, Gu and Yang obtained $\kappa =$ 103 Wm$^{-1}$K$^{-1}$ for a 1 μm-long MoS$_2$ at room temperature [35]. Their calculations also showed that only approximately 30 percent of the heat in MoS$_2$ is carried by ZA phonons, as opposed to 70-80 percent in graphene. They attribute the relatively smaller ZA phonon contribution to the thermal conductivity in MoS$_2$ to the larger scattering phase space for ZA phonons due to the absence of the plane-symmetry like in graphene.

The thermal transport properties of other TMDCs have not been studied much experimentally. Using the optothermal technique, the thermal conductivity of suspended CVD-grown monolayer WS$_2$ was measured Peimyoo and co-workers to be ∼32 Wm$^{-1}$K$^{-1}$ [36]. Interestingly, Gu and Yang predict that the thermal conductivity of monolayer WS$_2$ to be significantly larger than that of MoS$_2$ at $\kappa =$ 142 Wm$^{-1}$K$^{-1}$[35], which they attribute to the larger frequency gap between the acoustic and optical phonon branches in WS$_2$.

### 2.3.4 Black phosphorus

Single-layer BP, also called phosphorene, is the single-layer form of the thermodynamically stable allotrope black phosphorus and has a honeycomb-like puckered lattice of P atoms with an orthorhombic crystal structure. As a consequence of the puckering in BP, there are two different high-symmetry directions: The first corresponds to the direction along which the side view of the lattice shows an armchair-like atomic configuration and is denoted as the "armchair" direction while the second, denoted as the "zigzag" direction, is parallel to the zigzag-like edge.

Given its puckered structure, single-layer BP is much more ductile in the armchair direction than in the zigzag direction, with the Young's modulus estimated to equal 21.9 Nm$^{-1}$ and 56.3 Nm$^{-1}$ in the armchair and zigzag direction, respectively [37]. This anisotropy also suggests that the phonon group velocities and hence the thermal conductivity in the zigzag direction ($\kappa_{ZZ}$) should be higher than that in the armchair direction ($\kappa_{AC}$) [38,39]. Theoretical estimates of the room-temperature thermal conductivity of single-layer BP show that there is significant anisotropy, with the ratio $\kappa_{ZZ}/\kappa_{AC}$ varying between 2.2 and 5.5, although the absolute values of $\kappa_{ZZ}$ and $\kappa_{AC}$ at room temperature still vary widely [39–44] with $\kappa_{AC}$ varying between 4.59 and 36 Wm$^{-1}$K$^{-1}$ and $\kappa_{ZZ}$ varying between 15.33 and 136.2 Wm$^{-1}$K$^{-1}$. Given that these values are typically obtained using the Boltzmann Transport Equations combined with phonon lifetime estimates based on *ab initio* calculations [45], the variation in predicted thermal conductivity values has been attributed to differences in the details of the *ab initio* calculations [10].

Experimentally, there has been no characterization of the thermal conductivity in single or few-layer BP owing to the difficulty of isolating such samples. Thus, measurements of the thermal conductivity have been limited to BP flakes of at least 10 nm in thickness. Using micro-Raman spectroscopy, Luo and coworkers determined the armchair and zigzag thermal conductivities to be ~20 and ~40 Wm$^{-1}$K$^{-1}$ for BP films thicker than 15 nm, respectively, corresponding to a $\kappa_{ZZ}/\kappa_{AC}$ ratio of ~2 [46]. Their theoretical modeling also suggests that the observed thermal conductivity anisotropy is due mainly to the anisotropic phonon dispersion. Using the time-domain thermal reflectance technique, Jang and coworkers measured the thermal conductivity of relatively thick BP flakes, with thickness ranging between 138 and 552 nm, and obtained the maximum thermal conductivities of $\kappa_{ZZ} = 86 \pm 8$ Wm$^{-1}$K$^{-1}$ and $\kappa_{AC} = 34 \pm 4$ Wm$^{-1}$K$^{-1}$ [47]. In addition, they also determined the cross-plane component of the thermal conductivity to be $4.0 \pm 0.5$ Wm$^{-1}$K$^{-1}$. It should be noted that even though these experimentally measured thermal conductivity values are for relatively thick many-layer BP, they fall within the rather wide range of theoretically computed values for single-layer BP, suggesting the thermal conductivity of single or few-layer BP should be comparable.

### 2.4 Cross-plane thermal transport

In addition to in-plane heat conduction, another channel of heat dissipation is heat transfer at the interface of the 2D crystal in the cross-plane direction and it is characterized by its thermal boundary conductance (TBC) $G$ which determines the ratio of the interfacial heat flux to the temperature difference. For insulating or semiconducting 2D crystals, it is posited that interfacial heat transfer can occur via electron-phonon remote scattering [48], near-field radiation [49,50] and lattice (phonon) coupling [51]. However, the dominant heat transfer mechanism is through the lattice because heat transfer through electron-phonon scattering and near-field radiation is limited by the very low intrinsic charge carrier densities in 2D crystals.

In spite of the importance of cross-plane thermal transport for energy dissipation in van der Waals heterostructure, the mechanism of phonon-mediated interlayer heat transfer remains poorly understood partly because of the conceptual difficulty of applying traditional theories of interfacial heat transport such as the acoustic and diffuse mismatch models (AMM and DMM) [52], which have been widely used in the literature [53–56] and assume the existence of an extended volume on either side of the interface. For a single or few-layer 2D crystal supported on a substrate, the dimensionality mismatch and limited thickness of the 2D crystal make it hard to justify rigorously the application of the mismatch models. Moreover, it is unclear in such models what role the structure of the interface (e.g. interfacial bonding) plays in interfacial heat transfer.

As an alternative, an elasticity-based model describing the coupling between the flexural phonons of the graphene and its substrate has been proposed by Persson *et al* [51,57] and extended by Ong *et al* to encapsulated and multilayer 2D crystals [58,59]. In this model, heat is transferred between the flexural phonons of the 2D crystal and its substrate via the interfacial stresses and the rate of heat conduction depends on the elastic parameters of the 2D crystals and the substrate. We recall that it is argued [20] that for symmetry reasons, the flexural phonons are also the main heat carriers in in-plane thermal transport in freestanding graphene and that in supported graphene, its role is severely diminished as a result of interaction with the substrate [18]. However, this interaction with the substrate is not necessarily disadvantageous from a heat dissipation standpoint as it furnishes an alternative pathway for heat dissipation via cross-plane thermal transport.

In lieu of analytical models of cross-plane heat transfer, a popular computational approach to study the thermal boundary conductance of 2D crystals is the atomistic Green's function (AGF) method [60–62] which simulates the ballistic transmission of phonons across the interface. Another common approach is the use of molecular dynamics (MD) simulations [63–66]. These computational methods allow us to model interfacial heat transfer with atomistic fidelity and investigate the possible effects of material structure (such as grain boundaries) on the TBC [67] as well as to analyze the role of specific phonons and local thermal transport phenomenon. By analyzing the differences in the phonon lifetimes between supported and freestanding graphene, Ong and coworkers concluded that interfacial thermal transport is mainly mediated by flexural phonons [65]. Using equilibrium MD simulations, Ni and coworkers found that the interlayer thermal resistance in multilayer graphene to be around ~$10^{-9}$ $m^2KW^{-1}$ (or an interlayer thermal conductance of ~1 $GWm^{-2}K^{-1}$) [63].

Experimentally, there are relatively few studies of the thermal boundary conductance between the 2D material and its substrate, with most of them carried out for graphene. The most common techniques used are the 3ω method [53], time-domain thermoreflectance (TDTR) [54,68,69], and Raman thermometry-based methods [17,62]. The typically measured room-temperature TBC between graphene and its substrate is between 20 and 100 $MWm^{-2}K^{-1}$, about an order magnitude lower than the interlayer thermal conductance in multilayer graphene. On the other hand, the measured TBC between $MoS_2$ and its substrate is considerably smaller. Taube and coworkers estimated the TBC between $MoS_2$ and its $SiO_2$/Si substrate to be ~1.94 $MWm^{-2}K^{-1}$ [33] while Yalon *et al* obtained a much higher value of ~14

MWm$^{-2}$K$^{-1}$ [5,70] which is nonetheless still significantly lower than the typical TBC for graphene. Using a TDTR-based approach, Liu and coworkers showed that the TBC for the interface between MoS$_2$ and thin metal films is ~26 MWm$^{-2}$K$^{-1}$, significantly lower than the TBC between graphite and sputtered Al thin film (~73 MWm$^{-2}$K$^{-1}$) [71]. The same trend was observed in Ref. [62], where the TBC between graphene and h-BN (~52 MWm$^{-2}$K$^{-1}$) is about 3 times that of the TBC between MoS$_2$ and h-BN (~17 MWm$^{-2}$K$^{-1}$), and is attributed to its lighter C atoms and lower areal mass density which result in more phonon frequencies available in graphene for phonon transmission to the substrate. These differences in TBC show that heat dissipation across the interface depends on the native phononic character of the 2D material as well as the interactions at the interface.

## 3. Energy dissipation of 2D electronic devices in a steady-state

### 3.1 Energy dissipation in graphene and MoS$_2$ device

While the superior electronic [72] and thermal [16] properties of graphene provide a platform for testing solid state devices such as high speed field-effect transistors (FETs) [73], the intrinsic properties of the graphene are significantly affected by its environment. The electrons and holes traveling in a graphene prepared on SiO$_2$ substrate are scattered by polar phonons in SiO$_2$, leading to an order-of-magnitude reduction in the carrier mobility. In the case of thermal transport, the out-of-plane (flexural) phonons in graphene leak into its substrate through van der Waals interaction, resulting in an order-of-magnitude reduction of the thermal conductivity [18]. For future energy-efficient graphene-based electronics on substrates, the understanding of both electronic and thermal behaviors of graphene under operating conditions is needed because the operating power generates Joule self-heating in graphene, which changes its electronic and thermal transport characteristics [74]. To study energy dissipation in a graphene field-effect transistor (GFET), we need to know the temperature profile along the operating graphene channel, which can be obtained using various methods such as electro-Raman measurement [75], grey-body radiation [74,76–78] and scanning thermal microscopy [79,80].

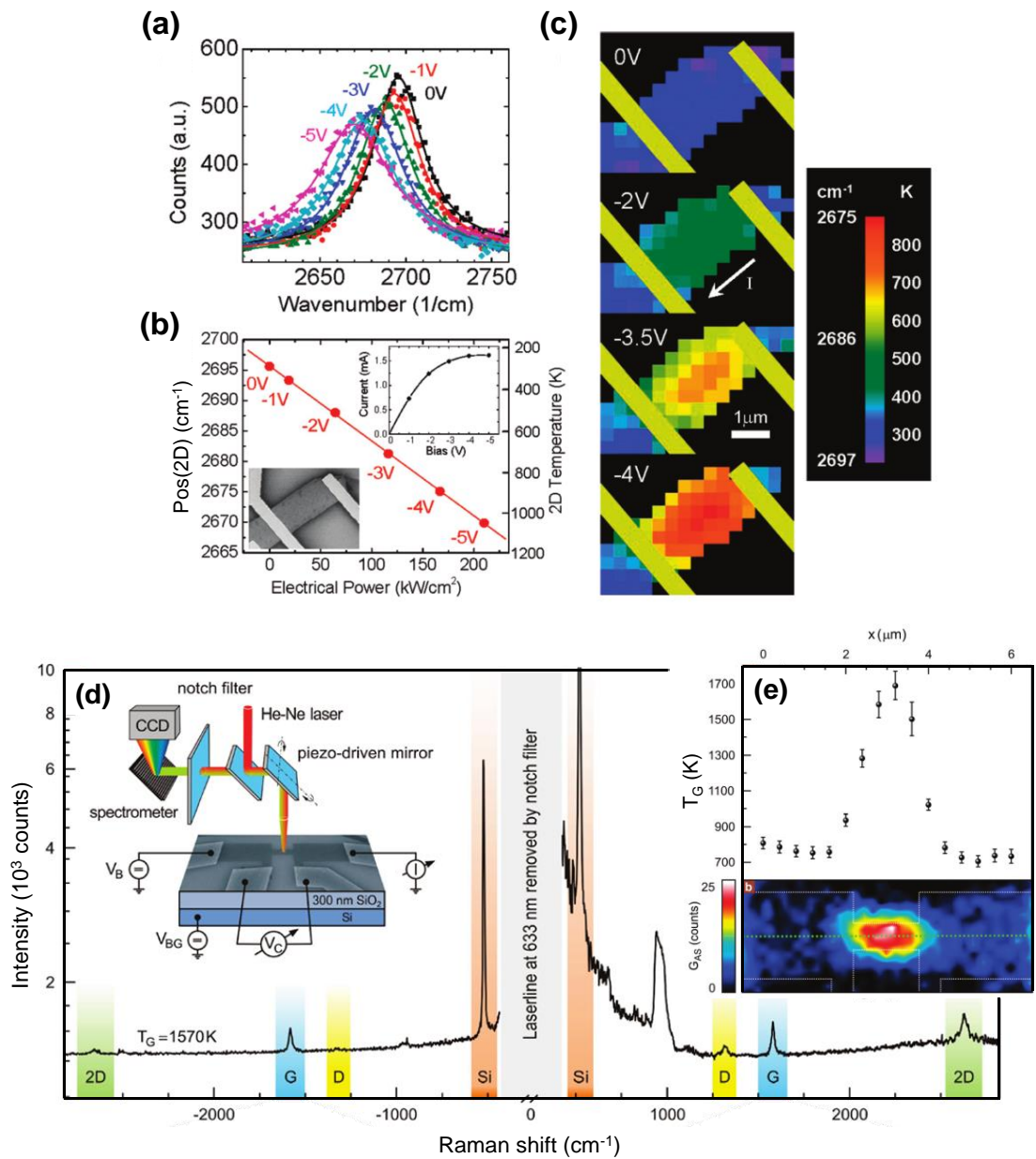

Figure 2. Raman thermometry. (a−c) Softening of the 2D band of graphene with increasing bias voltage and the temperature distribution in the graphene channel at different bias-voltage values. (d),(e) Anti-Stokes and Stokes spectrum of graphene under bias voltage and temperature distribution in the graphene constriction extracted from the ratio between the anti-Stokes and Stokes OP mode intensities. Figures (a−c) are reproduced with permission from [75], © 2009 American Chemical Society; (d) and (e) from [81], © 2010 American Chemical Society.

The electro-Raman method utilizes two different thermometric mechanisms: i) the softening of the optical phonon (OP) mode with increasing baseline temperature [75] and ii) the ratio between the anti-Stokes and Stokes OP mode intensities [81]. The former and latter techniques are sensitive to the acoustic phonon (AP) and OP temperatures, respectively. In these techniques, the laser power should be low enough to avoid laser-induced heating.

In thermometry based on OP mode softening, the 2D band near 2700 cm$^{-1}$ shows a red shift that varies linearly with temperature. For example, in Ref. [75], there is a shift of −29.4 Kcm$^{-1}$ as the baseline temperature, $T_b$ increases from 300 to 1000 K, as shown in figure 2(a−c). The 2D band peak also exhibits a red shift with increasing electrical power at $T_b = 300$ K, which provides information related to the electric power dissipation within the graphene channel. The authors found that the graphene temperature can reach ~1000 K at an input power density of 210 kWcm$^{-2}$ when the device is placed within an environment with dry nitrogen flowing to prohibit oxidation. Their analysis showed that ~80 % of the generated heat is directly dissipated to the SiO$_2$ layer underneath the graphene channel and ~20 % to the metal electrodes and graphene extended beyond the device. It was suggested that that the heat could also dissipate into the substrate via remote scattering of the hot graphene electrons by the polar surface phonons in SiO$_2$ layer, resulting in a much lower graphene/SiO$_2$ TBC value of ~25 MWm$^{-2}$K$^{-1}$ [82].

Alternatively, the temperature of the G-band OPs ($T_G$) in a graphene can be directly determined from the ratio between the Stokes ($I_S$) and anti-Stokes ($I_{AS}$) G band intensities: $I_{AS}/I_S$ = $[(\omega_L+\omega_G)/(\omega_L-\omega_G)]^4 \exp(-\hbar\omega_G/k_B T_G)$. Here, $\omega_L/\hbar$ and $\omega_G/\hbar$ are the frequency of the laser and the G-phonon, and $\hbar$ and $k_B$ are Planck and Boltzmann constants, respectively. Using this method, Chae *et al* [81] found that the temperature was raised up to ~1600 K under a power density of 270 kWcm$^{-2}$ in a graphene constriction region [length ($L$) = 1.5 µm and width ($W$) = 0.6 µm] (figure 2(d) and (e)). Their analysis suggests that the strong electron-phonon coupling near the Dirac point can lead to a high non-equilibrium G-band OP population.

Based on this study, one can expect the hot electrons and the OPs to have a similar temperature given this strong electron-OP coupling. Indeed, it has been shown in *in-situ* measurements of the grey-body radiation combined with Raman spectroscopy under high electrical bias that the excited G-OPs are in equilibrium with the electrons at least up to ~1500 K [77]. Here, the electron and G-OP temperatures are given by $T_{el}$ and $T_{OP}$, respectively. The spectral radiance, $u$ is determined by Planck's law: $u = \varepsilon(2E^3/h^2c^2)[\exp(E/k_B T_{el}) -1]^{-1}$ for grey-body radiation. Here, $c$ and $E$ are the speed of light in vacuum and energy of the emitted photons, respectively, while $\varepsilon$ is the emissivity, a material property that reflects how the material emits photons efficiently. For monolayer graphene, one can use $\varepsilon$ = 2.3 % [83]. By analyzing the softening of the G band, it has been shown that the OPs (or hot electrons) are not in full equilibrium with the lower-energy APs because of the strong electron-phonon interaction which can explain the current saturation behavior in graphene field-effect transistors with the surface polar phonons in the SiO$_2$ layer also playing a role [84]. On the other hand, the equilibrium between the electrons, OPs and APs can be sustained up to at least 700 K [76].

We now turn our attention to the spatial dependence of energy dissipation along the graphene channel under high electrical bias. Like in local Raman spectroscopy, the temperature distribution along the graphene can be inferred from local thermal infrared emission measurements along points in the graphene channel. The infrared scope takes the temperature profile of the sample at one time. The spectral range of detected photons is determined by the infrared sensor. For example, the liquid nitrogen-cooled InSb detector can detect IR photons in the 2-4 μm wavelength range. To get enough photons for temperature analysis, the temperature of a sample should be higher than ~70 K. Using this method, Bae *et al* [74] have shown that a hot spot can form in the graphene channel and move along the graphene as a the gate voltage is varied (figures 3(a-d)). This is because the charge neutral point which corresponds to the local maximum resistance, and hence the local hot spot can move along the channel, with its position depending on the gate voltage-induced formation of a local ambipolar region in the diffusive regime.

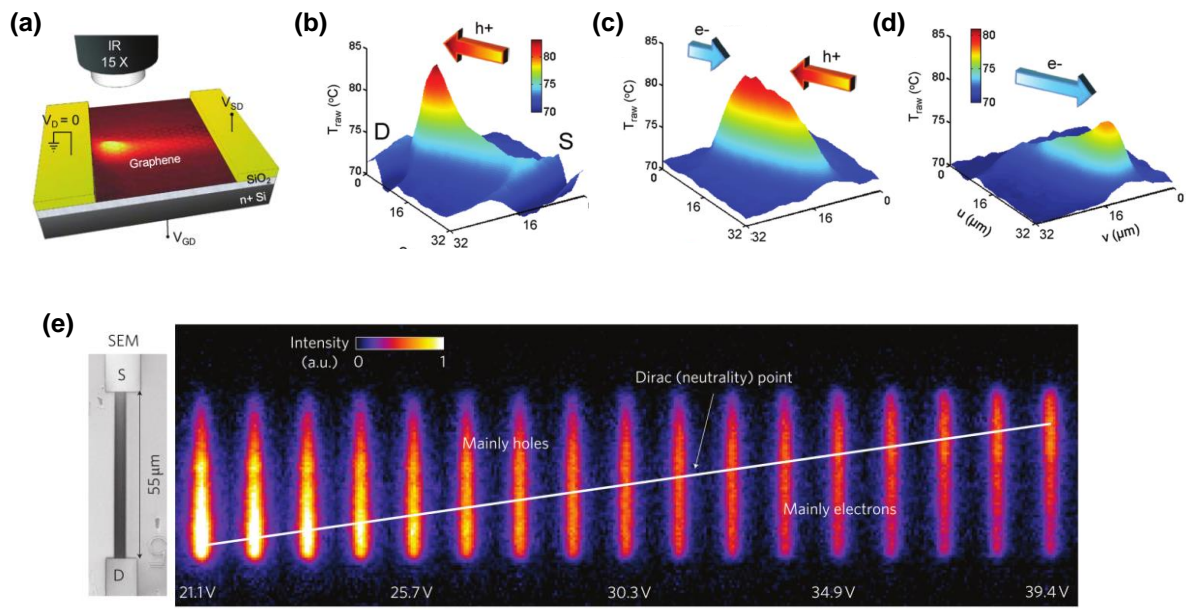

Figure 3. Spatial dependence of energy dissipation. (a−d) Infrared-scope thermal imaging results. The hot spot moves along the graphene channel when the carrier density is varied with bias and gate voltages. (e) Grey-body radiation intensity along the graphene channel with changing gate voltage, where a bright color indicates a relatively hot region. Figures (a−d) are reproduced with permission from [74], © 2010 American Chemical Society; (e) from [76], © 2010 Macmillan Publishers.

The same phenomenon can also be found in bilayer graphene. One should note that the photons detected by the thermal imaging originate from the top of the Si layer and not the graphene because the graphene and SiO$_2$ layer are nearly transparent to photons in the 2-4 μm wavelength range. Therefore, to extract the real temperature of the graphene, one needs a conversion process as described in Ref. [74]. Freitag *et al* [76] also arrived at the same conclusion with an infrared analysis, where the infrared spectrum from the sample was detected by a liquid nitrogen-cooled HgCdTe (λ = 1.5 μm-3 μm) detector array (figure 3(e)). However, because of the low spatial resolution limit (e.g., 2.6 μm for Ref. [74]), the efficient application of this infrared thermal imaging technique is limited to graphene samples with a channel longer than 15 μm.

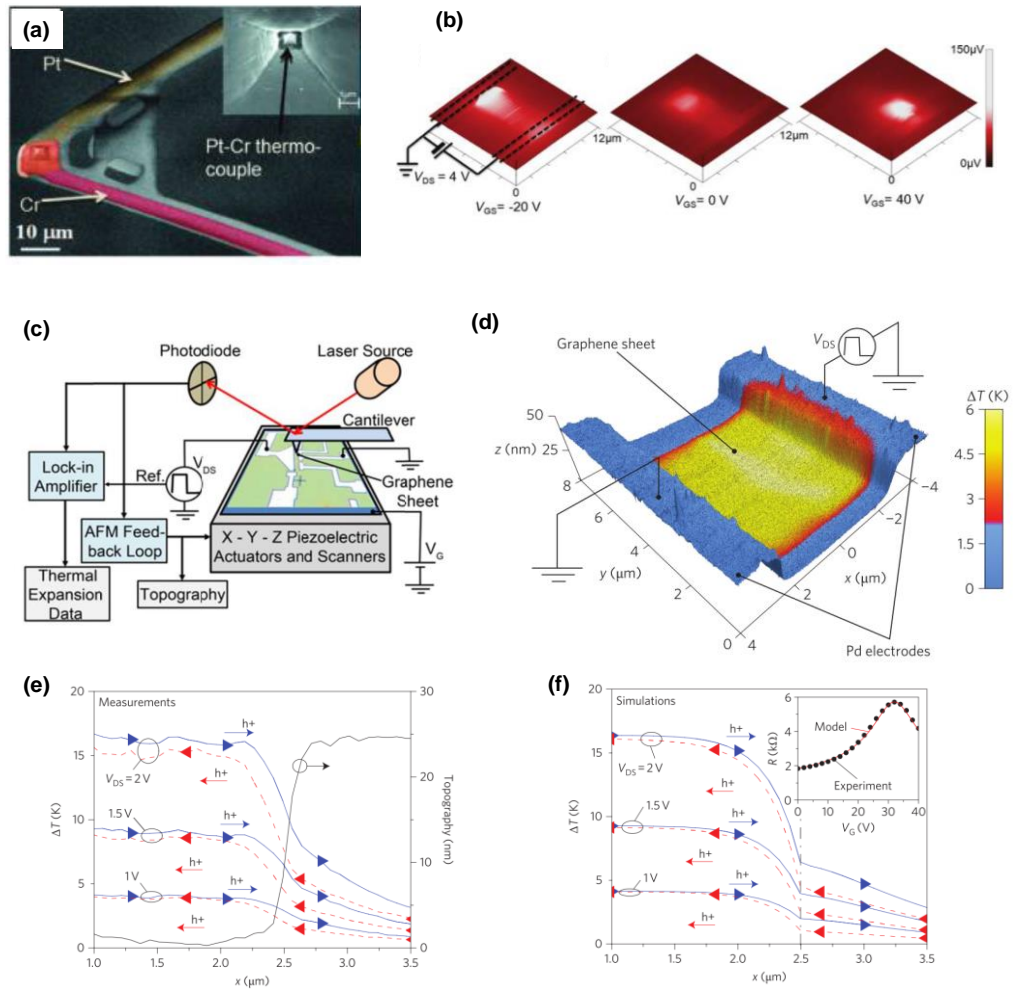

Figure 4. Scanning thermal microscopy (SThM). (a),(b) Scanning electron microscope image of SThM probe and thermal mapping results with a bias voltage of 4 V at different gate voltages. (c),(d) Setup for the scanning Joule expansion microscopy and temperature distribution with a bias voltage of 1.5 V. (e), (f) Measured and simulated temperature profiles at the graphene/metal contact depending on current direction. Figures (a),(b) are reproduced with permission from [80], © 2011 American Chemical Society; (c−f) from [79], © 2011 Macmillan Publishers.

To overcome the low spatial resolution problem, scanning thermal microscopy techniques with a cantilever based on atomic force microscopy have been developed. These methods can measure the temperatures of the APs. The first example for that is based on a microscale Pt-Cr thermocouple attached to the tip of the cantilever and was demonstrated by Jo *et al* [80] who successfully performed the thermal scanning of a ~7 μm long graphene channel with ~100 nm spatial resolution and confirmed earlier results measured with the infrared scope (figures 4(a) and (b)). Although the thermocouple tip reads voltage change due to the temperature change, it is hard to get the real temperature of the sample because the convection of the air around the tip affects the temperature reading [85].

Another interesting thermal scanning technique is the scanning Joule expansion microscopy [86]. To perform the experiment, the graphene sample is covered by a polymer such as a PMMA. Because the polymer expands with temperature, if one measures the height increase for a given bias condition, the temperature can be estimated from the changing height while taking into account the thermal environmental. For this technique, the spatial resolution can reach ~10 nm. Using this technique, Grosses *et al* [79] found that the graphene/metal contact region can exhibit the Peltier effect as well as Joule heating and current crowding. For instance, a graphene/metal junction can be cooled or heated, depending on the current flow direction (figures 4(c−f)).

The greatly elevated temperature of a graphene channel undergoing Joule-heating can induce breakdown. In air, the burning temperature of carbon nanotubes (CNTs) is ~600 °C. Given that carbon atoms in both of CNT and graphene are connected to neighboring carbon atoms by $sp^2$ bonds in a hexagonal crystal lattice, we speculate that the breakdown temperature of graphene is also ~600 °C. Assuming this breakdown condition, Liao *et al* [87] investigated graphene nanoribbons (GNRs) on a $SiO_2$ layer under high electrical bias conditions, up to breakdown. They found that the current density reached to $4\times10^8$ A/cm$^2$ in ~15 nm narrow GRNs, which is larger than that of general micron-sized graphene channels. This is because heat from the GNRs is spread over an effectively larger volume of $SiO_2$. Importantly, the breakdown and corresponding high field behavior of GNRs yield information of the thermal conductivity of the GNRs, which was determined to be ~80 Wm$^{-1}$K$^{-1}$ at room temperature (figure 5(a)).

For suspended graphene, the temperature of the graphene can be easily increased because there is no substrate acting as a heat sink. Dorgan *et al* [88] showed that a few micron-sized suspended graphene in vacuum can reach high temperatures over 1000 K at a high field near 1 V/µm. Based on the analysis of the breakdown voltage and temperature, the thermal conductivity at 1000 K was estimated to be as much as 300 Wm$^{-1}$K$^{-1}$, corresponding to 2500 Wm$^{-1}$K$^{-1}$ at room temperature (figure 5(b)). This value is similar to recent thermometry results for suspended graphene [89].

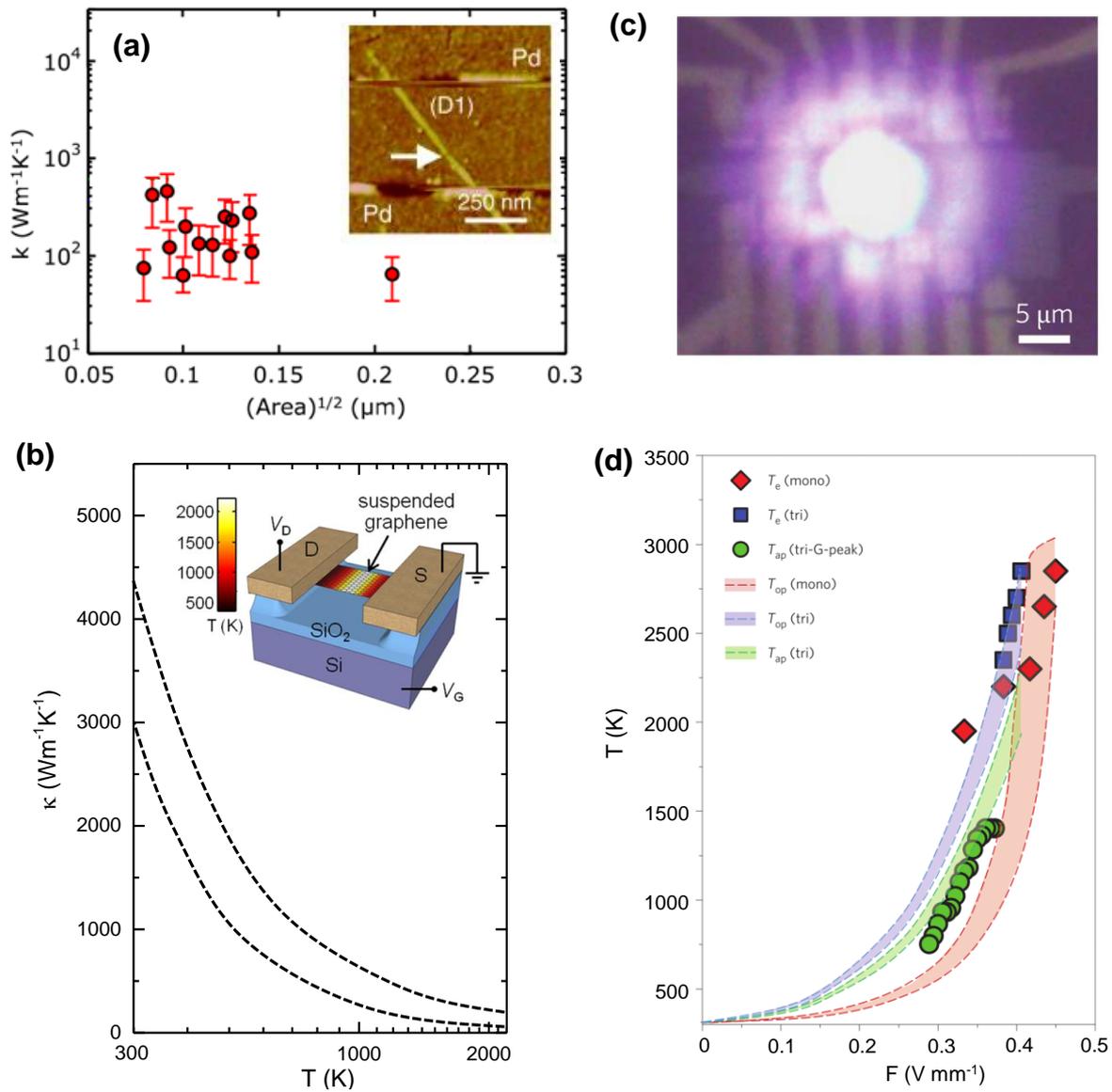

Figure 5. Joule-heated breakdown of graphene. (a) Thermal conductivity of graphene nanoribbons estimated based on the Joule-heated breakdown event at $T_{BD} \approx 600$ K in air. The inset: Atomic force microscopy image of a graphene nanoribbon after breakdown, as indicated by an arrow. (b) Suspended graphene thermal conductivity (upper and lower dashed lines: upper and lower bounds, respectively) estimated from the breakdown event at $T_{BD} \approx 2200$ K in vacuum. (c),(d) Optical microscope image of visible light emission from a suspended graphene and measured (scattered symbols) and estimated (shaded regions) temperatures as a function of the biased electric field ($F$) in the suspended graphene. Figure (a) is reproduced with permission from [87], © 2011 American Physical Society; (b) from [88], © 2013 American Chemical Society. (c) and (d) from [90], © 2015 Macmillan Publishers.

Since the electrically biased suspended graphene device has no vertical heat dissipation and heat transport along the suspended graphene channel is strongly degraded by the serious self-Joule heating, the electron temperature at the center can reach ~2800 K, which is impossible for graphene on SiO$_2$ because of rapid cooling of hot carriers through remote scattering. Recently, it was demonstrated that the hot electrons at the center of the graphene can thermally emit visible light of which the color is tunable from red to yellow through the interference of the emitted light between the suspended graphene and the substrate (figures 5(c) and (d)) [90].

In TMDC 2D layered semiconductors, it is expected that the thermal interface between the semiconductor and the substrate plays a role in the heat dissipation given the 2D nature of the system. Based on Raman spectroscopic characterization of MoS$_2$ and SiO$_2$ in an operating MoS$_2$ device on a SiO$_2$ substrate, the TBC at MoS$_2$/SiO$_2$ was determined to be ~14 MWm$^{-2}$K$^{-1}$ at room temperature as aforementioned, which is near the very low end of TBC values for known solid-solid interfaces [5]. This indicates that the energy dissipation in 2D semiconductors is seriously limited by the interface with the substrate. A. Behranginia *et al* [91] found that the TBC between ~7 nm thick WSe$_2$ and SiO$_2$ was between 10-32 MWm$^{-2}$K$^{-1}$ by analyzing how the low-frequency shear mode depends on the temperature of a WSe$_2$ device, which experiences a breakdown at ~370 °C for a power density of ~1.2 mW/μm$^2$. On the other hand, experiments performed on MoS$_2$ and MoSe$_2$ (single and bilayers) flakes suspended over holes defined in a SiO$_2$/Si substrate using the optothermal Raman technique determined the TBC to be ~0.5 MWm$^{-2}$K$^{-1}$ and ~0.1 MWm$^{-2}$K$^{-1}$, respectively [92]. This indicates that the TBC can vary depending on the interface between the TMDCs and substrate.

## 3.2 Energy dissipation in van der Waals heterostructures

The cooling mechanism of hot carriers by remote scattering between the graphene and its substrate can be tuned by changing the substrate. In the remote scattering, the polar OP energy determines the energy of mobile charge carriers in graphene and the maximum electron temperature [93]. For instance, the maximum temperature of hot carriers in graphene on $SiO_2$ is limited to ~1100 K which corresponds to the $SiO_2$ polar phonon energy of 60-80 meV. If *h*-BN is used the substrate, it is expected that the carrier temperature can reach up to ~2000 K because of its polar phonon energy of 150-200 meV. Recently, it was shown that an h-BN/graphene/*h*-BN vdW heterostructure thermally emits a visible light when the electron temperature reaches 2000 K (figures 6(a) and (b)) [94]. The efficient electronic cooling via near-field coupling to hybrid polaritonic modes in the *h*-BN substrate [95] allowed a high-speed graphene light emitter with ~10 GHz modulation.

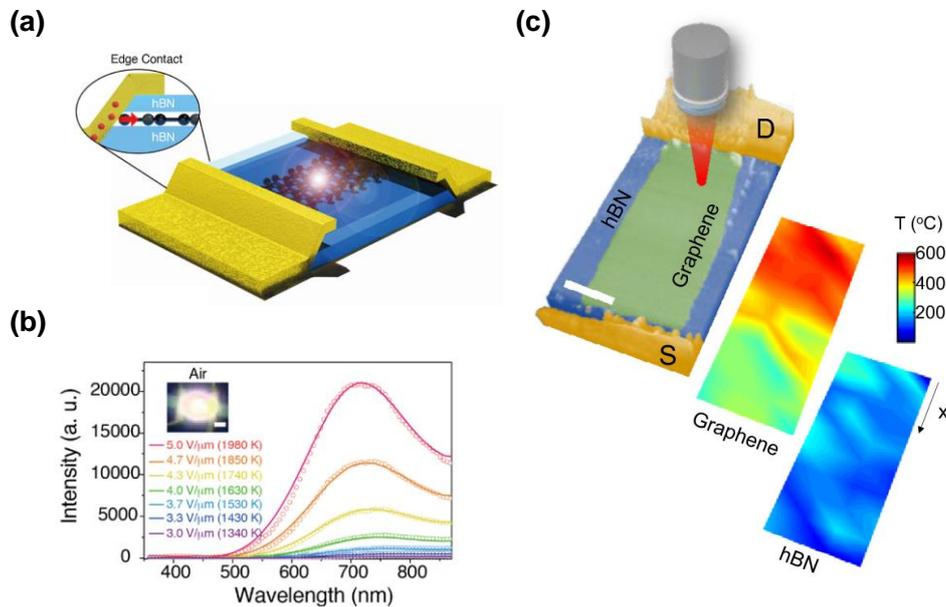

Figure 6. Energy dissipation in graphene on *h*-BN. (a) *h*-BN encapsulated graphene thermal light emitter. (b) Radiation spectrum (scattered symbols) from the graphene and calculation results (solid lines) based on electron temperature and grey-body radiation for various bias electric field (*F*). Inset: optical image of light emission from the device at $F$ = 4.3 V/μm (scale bar: 6 μm). (c) Thermography of graphene/*h*-

BN device. Top: Atomic force microscope image of the device (scale bar: 1 μm). Middle and bottom: temperature distribution in graphene and *h*-BN substrate at bias voltage of 8.5 V. Figures (a) and (b) are reproduced with permission from [94], © 2018 American Chemical Society; (c) from [96], © 2018 IOP Publishing Ltd.

A spatial heat dissipation study of graphene/*h*-BN devices was performed using spatially resolved Raman spectroscopy in air [96] showed that a high electric-field-induced doping level significantly changes the mechanism of heat dissipation into the *h*-BN, especially near the breakdown temperature of graphene, ~600 °C (figure 6(c)).

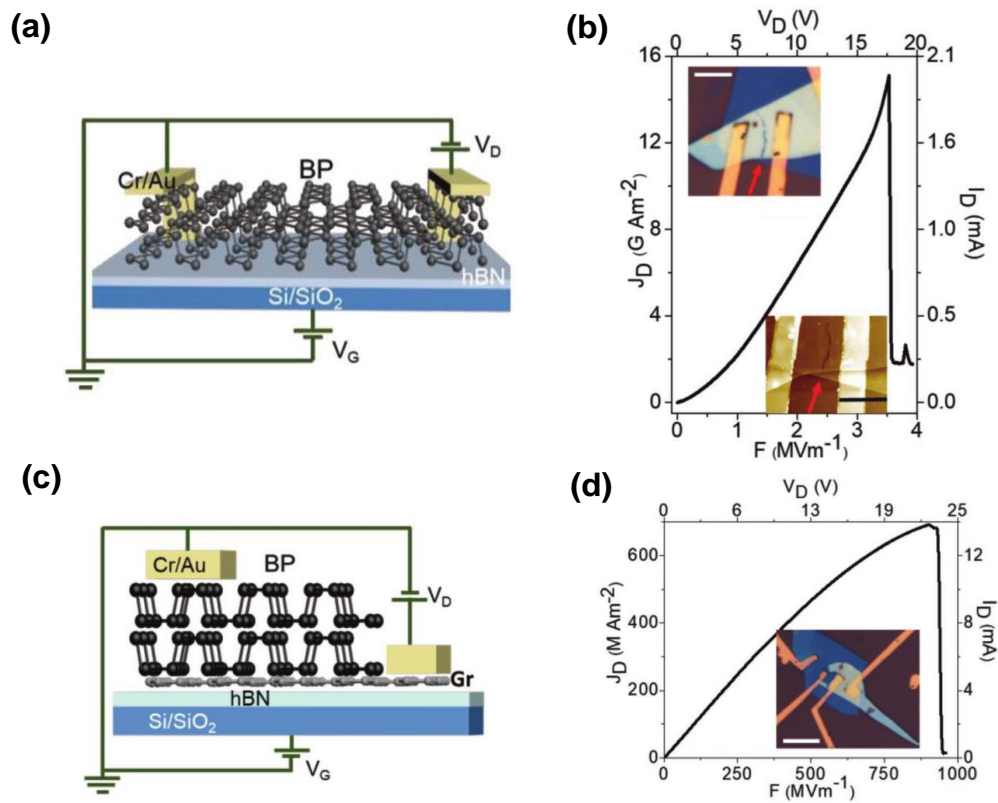

Figure 7. Energy dissipation in a BP heterostructure. (a) Schematic of BP/*h*-BN lateral FET. (b) Current (density) as a function of applied electric field. Top and bottom insets: optical and atomic force microscope images of the device after breakdown at $F \approx 4$ MVm$^{-1}$, respectively (scale bar: 5 μm). (c) Schematic of vertical BP/graphene device on *h*-BN. (d) Current (density) as a function of applied electric field. Inset: optical image of device after breakdown at $F \approx 900$ MVm$^{-1}$ (scale bar: 5 μm). Figures are reproduced with permission from [97], © 2018 WILEY-VCH Verlag Gmbh & Co.

The vdW heterostructure configuration is useful for enhancing hot carrier cooling in 2D semiconductors. For example, it was shown that a BP FET undergoes a Joule-breakdown under a moderate electric field of ~4 V/μm due to inefficient energy dissipation given the relatively low thermal conductivity of BP of 20-40 Wm$^{-1}$K$^{-1}$ (figures 7(a) and (b)) [97]. A BP/graphene/hBN structure however showed an enhanced tolerance of an up to two orders of magnitude stronger field than that of a BP device, because the relatively high thermal conductivity of the graphene and hBN provide an efficient cooling (figures 7(c) and (d)).

## 4. Dynamics of thermalization and cooling processes of hot electrons in graphene and van der Waals heterostructures

### 4.1 Thermalization of photoexcited electrons in graphene materials

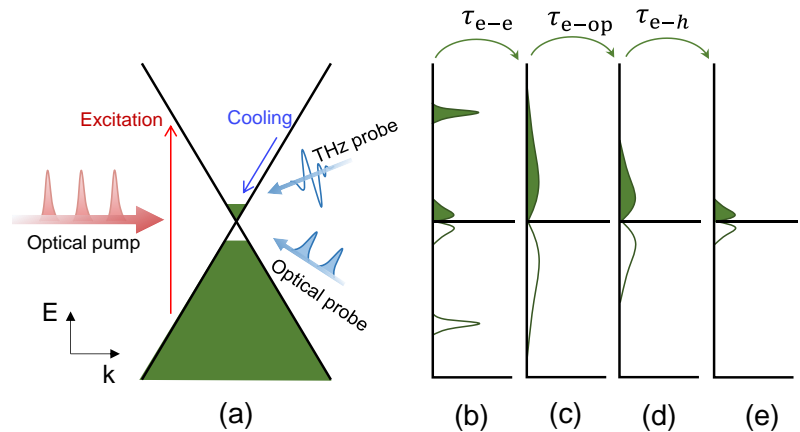

Figure 8. Time-resolved optical pump-(optical or terahertz) probe spectroscopy. (a) Electronic band of graphene with intrinsic electron and hole populations near the Dirac point. The optical pump pulses generate photoexcited carriers and a relatively weak probe pulse is used to measure the transmittivity change through the sample at various time delays of the optical probe pulses with respect to the pump pulses. In the case of a terahertz probe, the optical pump is delayed with respect to the terahertz probe. (b−e) A schematic of the cooling process of optically excited hot electrons. Figures are reproduced with permission from [98] and [99], © 2008 American Institute of Physics and © 2008 American Chemical

Society.

The pulsed pump laser having energy of ~1.6 eV excites electrons from the valence to the conduction band, creating a population of photoexcited electrons in the conduction band, as shown in figure 8(a), with the photoexcited carrier density estimated to be in the $10^{11} - 10^{12}$ cm$^{-2}$ range [98]. As a result of hot-electron inelastic intraband scattering from the electron-electron interaction, the photoexcited carriers equilibrate rapidly with the original carrier population [100], leading to a final Fermi-Dirac-like distribution [figure 8(c)] that is at a much higher temperature than the lattice temperature. The relaxation time ($\tau_{e-e}$) for this equilibrium process is only 10 – 50 fs (figure 8(b)). After this fast relaxation, the subsequent cooling of the hot carriers takes place through the lattice via carrier-optical phonon (OP) intraband scattering in the 0.15 – 1 ps range ($\tau_{e-op}$) (figure 8(d)), followed by the electron-hole recombination in the 1 – 15 ps range ($\tau_{e-h}$) (figure 8(e)). Such short time scales of sub- and picoseconds have been measured by time-resolved optical pump-probe or optical-pump/THz-probe spectroscopy (see figure 8(a)).

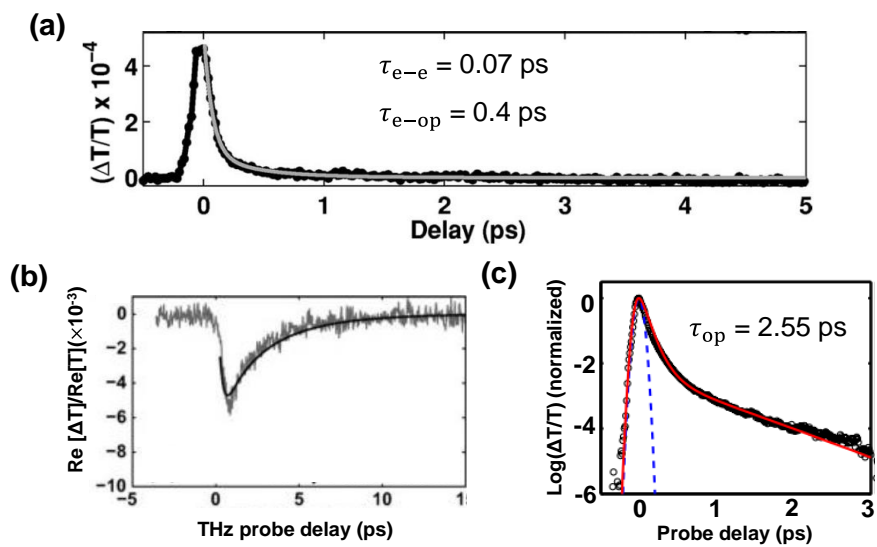

Figure 9. Transient transmittivity spectroscopy for an epitaxially grown graphene film measured by (a)

the optical pump-probe on a *6H*-SiC wafer to measure $\tau_{e-e}$ and $\tau_{e-op}$, (b) optical pump-probe terahertz spectroscopy spectroscopy for epitaxial graphene for $\tau_{e-h}$, and (c) the optical pump-probe spectroscopy for $\tau_{op}$. Figure (a) is reproduced with permission from [98], © 2008 American Institute of Physics; (b) from [99], © 2008 American Chemical Society; (c) from [101], © 2010 American Institute of Physics.

Figure 9(a) shows an example of the time dependent change in the transmittivity normalized to the transmittivity in the absence of the pump pulse (Δ*T*) for an epitaxially-grown few-layer graphene on a *6H*-SiC wafer, measured with the optical pump-probe spectroscopy [98]. After photoexcitation, the differential transmittivity decays exponentially with two time scales: a fast initial relaxation time of $\tau_1$ ~ 70 fs related to $\tau_{e-e}$, followed by a slower relaxation time of $\tau_2$ ~ 0.4 ps corresponding to $\tau_{e-op}$. For different graphene samples, the fast initial relaxation time was measured to be in the range of 70-120 fs, which is comparable to the pulse width of ~85 fs and thus of a significant uncertainty level. Subsequent optical pump-probe spectroscopy measurements with a 7-fs pulse width performed on freestanding thin graphite films revealed rapid intraband carrier equilibration within 30 fs [102]. The slower second relaxation time in Ref. [98] was in the 0.4 – 1.7 ps range, which is related to the carrier-OP scattering time and includes carrier recombination. Time-scale measurements based on the optical pump-probe spectroscopy to distinguish carrier-OP scattering and carrier recombination are difficult because optical pump-probe spectroscopy is sensitive to the interband conductivity of graphene and can only probe the time evolution of the carrier occupation at a specific energy window within the bands. The terahertz (THz) response in graphene at room temperature is sensitive to the intraband, total-carrier conductivity and carrier energy distribution [99]. Thus, the THz probe is useful for studying the carrier relaxation and recombination processes in graphene. Figure 9(b) shows the time evolution of the transmission which exhibits two distinct features: (i) an initial rapid decrease to ~1 ps, followed by (ii) a slow increase from 1 to 15 ps. The former indicates an increase in conductivity due to carrier cooling by Ops while the slow increase in the transmission is because of electron-hole recombination, resulting in the decrease of conductivity.

After the equilibration between the carriers and OPs within 1 ps, the cooling becomes limited by the slowing energy exchange between the carriers and OPs and the anharmonic decay of OPs. The OPs generated in graphene mainly decay into acoustic phonons (APs) through anharmonic scattering and lattice defects [101]. The anharmonic scattering-limited hot OP lifetime has been theoretically estimated to be in the 2–3 ps range for phonon temperatures in the 500-900 K range [103]. The OP lifetime has also been measured using optical pump-probe spectroscopy. The measured differential transmission transients show two distinct time scales, as shown in figure 9(c): (i) a rapid decrease to almost 10% of its peak value in 400-500 fs, and then (ii) a slow decrease. The hot photoexcited carriers dissipate most of their energy to the OPs in the first 0.5 ps after photoexcitation. This leads to the generation of hot OPs, resulting in a bottleneck in the subsequent cooling of the carriers. Because the relaxation dynamics of the carriers and the OPs are strongly coupled, the effects of electron-OP coupling should not be neglected in the interpretation of the experimental data. Thus, taking the intravalley or intervalley intraband electron-phonon scattering processes into account, the measured transient differential transmission was fitted to calculations of the electron and OP temperatures with a parameter, OP life time, $\tau_{op}$, which was estimated as 0.5~2.5 ps.

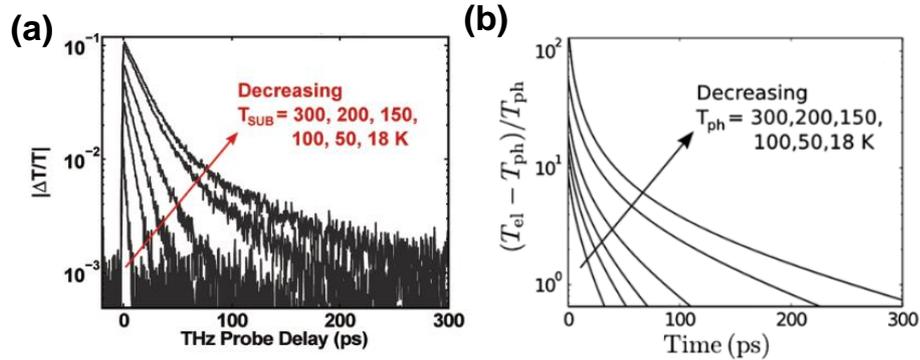

Figure 10. Supercollision scattering process. (a) Measured differential terahertz probe transmission for an epitaxial graphene sample at various temperatures. (b) Calculated results based on disorder-assisted electron-phonon scattering, which is consistent with (a). Figure (a) is reproduced with permission from [104], © 2011 American Chemical Society; (b) from [105], © 2012 American Physical Society.

On the other hand, it was found that the tail of the relaxation transients decay more gradually as temperature decreases, reaching several hundred picoseconds below 50 K, as shown in figure 10(a) [104]. It has been predicted that the cooling of hot carriers in graphene becomes very slow when the carrier energy is less than the OP energy [104] because carrier cooling in this case can occur only through AP emission. Since the OP energies of graphene are ~0.196 and 0.162 eV, the OP energy bottleneck becomes the main limiting factor for carrier cooling in graphene-based (opto)electronic devices. The very slow tails of the relaxation transients observed at low temperatures, where carrier temperature reaches 250 K, are attributed to the inefficient carrier relaxation from OP emission as the carrier energy distribution moves close enough to the Dirac point. Since the longitudinal AP scattering is also inefficient in cooling the carriers, the carrier cooling rate slows down further. Thus, the loss of heat from the graphene based on the AP scattering occurs over a time scale of 25–200 ps. Such inefficient cooling by the APs is due to the small Fermi surface and momentum conservation in the electron-AP scattering process [106–108]. In this case, the cooling power is given by $P_{AP} = A_1 T_{el}^4 (T_{el} - T_l)$ for $E_F \ll k_B T_{el}$ and $P_{AP} = A_2(T_{el} - T_l)$ for $E_F \gg k_B T_{el}$, where $T_{el}$ ($T_l$) is the electron (lattice) temperature, $E_F$ is the Fermi energy and $k_B$ is the Boltzmann constant. However, the temperature dependence of the tail of the differential transmittivity in figure 10(a) has also been theorized to be affected by disorder-assisted processes occurring through the so-called supercollisions, as shown in figure 10(b). Song *et al* [105] showed that the supercollision processes play an important role over a wide temperature range and up to room temperature. Since the energy transfer between electrons and Aps in graphene without disorder is limited by momentum conservation and the small Fermi surface in graphene, only a small part of $k_B T_{el}$ can be dissipated, for instance, $\Delta E_{AP}/k_B T \leq 2v_s/v_F \sim 0.04$. Here $\Delta E_{AP}(\leq 4$ meV) is the emitted AP energy and $v_F(v_s)$ is Fermi (sound) velocity [109]. This happens when $T_l$ is higher than the Bloch-Grüneisen temperature ($T_{BG} = (2v_s/v_F)E_F$), which is a characteristic temperature when the diameter of the Fermi surface equals the maximum phonon wave vector. On the other hand, in disorder-assisted cooling, the entire thermal distribution of APs can contribute to scattering, which leads to energy dissipation of order $k_B T$ per scattering event. Here, the exchange of momentum and energy between electron and phonon is mediated by the impurity scattering. The disorder-assisted cooling power is given by $P_{sc} = A(T_e^3 - T_l^3)$, where $A = 9.62 \frac{a^2 g^2(\mu) k_B^3}{\hbar k_F l}$, $a$ is the electron-phonon coupling, $g(\mu)$ is the density of states at the Fermi level per one spin or valley, $k_F$ is the Fermi wave vector, $l$ is the electron mean free path and $k_B T_{el(ph)} \ll \mu$. The more rapid cooling with increasing disorder is reflected by the disorder concentration term, $k_F l$ [98]. For $k_F l = 20$, for example, the disorder-assisted cooling can become up to 100 times more efficient compared to the disorder-free cooling. In the disorder-assisted cooling, the power-law cooling dynamics is given by $T_e(t) = \frac{T_{e,0}}{1+(A/\beta)(t-t_0)T_{e,0}}$, where $\beta = \frac{\pi^2}{3}Ng(\mu)k_B^2$ and $N$ is the spin and valley degeneracy.

**4.2 Supercollision cooling process in operating graphene devices**

Following our earlier introduction to the basic concepts about the thermalization and cooling processes of photoexcited electrons in graphene materials, we will discuss the cooling process of photoexcited electrons in graphene-based optoelectronic devices. In the devices, the photothermal effect (PTE) has been used to keep track of the hot-carrier cooling process near the Fermi level. Recently, the supercollision process was successfully demonstrated in graphene-optoelectronic devices using the PTE [109]. The photocurrent generated by the PTE is described by $I_{PTE} = (S_2 - S_1)\Delta T/R$, where $S_1$ and $S_2$ are the thermoelectric powers (or Seebeck coefficients) at two different regions 1 and 2, respectively, $\Delta T$ is the temperature difference between electron and electrode heat sink the two locations and $R$ is the resistance between the two locations [110,111]. In the Mott relation for a degenerate system, $S$ is proportional to the electron temperature and the thermoelectric current is given by the relation $I_{PTE}(t) = bT_e(t)(T_e(t) - T_l)$, where $b$ is the photocurrent proportionality constant related to the Seebeck coefficient. Within the supercollision framework, we have $I_{PTE} = b(P_{in}/A)^{2/3}$ for $T_e \gg T_l$ and $I_{PTE} \approx bP_{in}/3AT_l$ for $T_e - T_l \ll T_l$ at steady-state condition ($P_{sc} = P_{in}$) such as continuous laser irradiation. In the case of a pulsed laser, the electrons are heated to $T_0$ after the initial OP emission process, followed by the disorder-assisted cooling dynamic process. In that case, an average current becomes proportional to $F_{in}^{0.5}$, where $F_{in}$ is the effective incident laser pulse energy. These predictions for continuous and pulsed laser irradiations were experimentally confirmed as shown in figures 11(a−c). In figure 11(d), the transient photocurrent (TPC) shows a $1/t_d$ thermal decay, predicted by the supercollision model [105]. It was shown that the disorder-assisted cooling process plays a role for a broad temperature range of $T_e$ (20−3000 K) and $T_l$ (10−295 K).

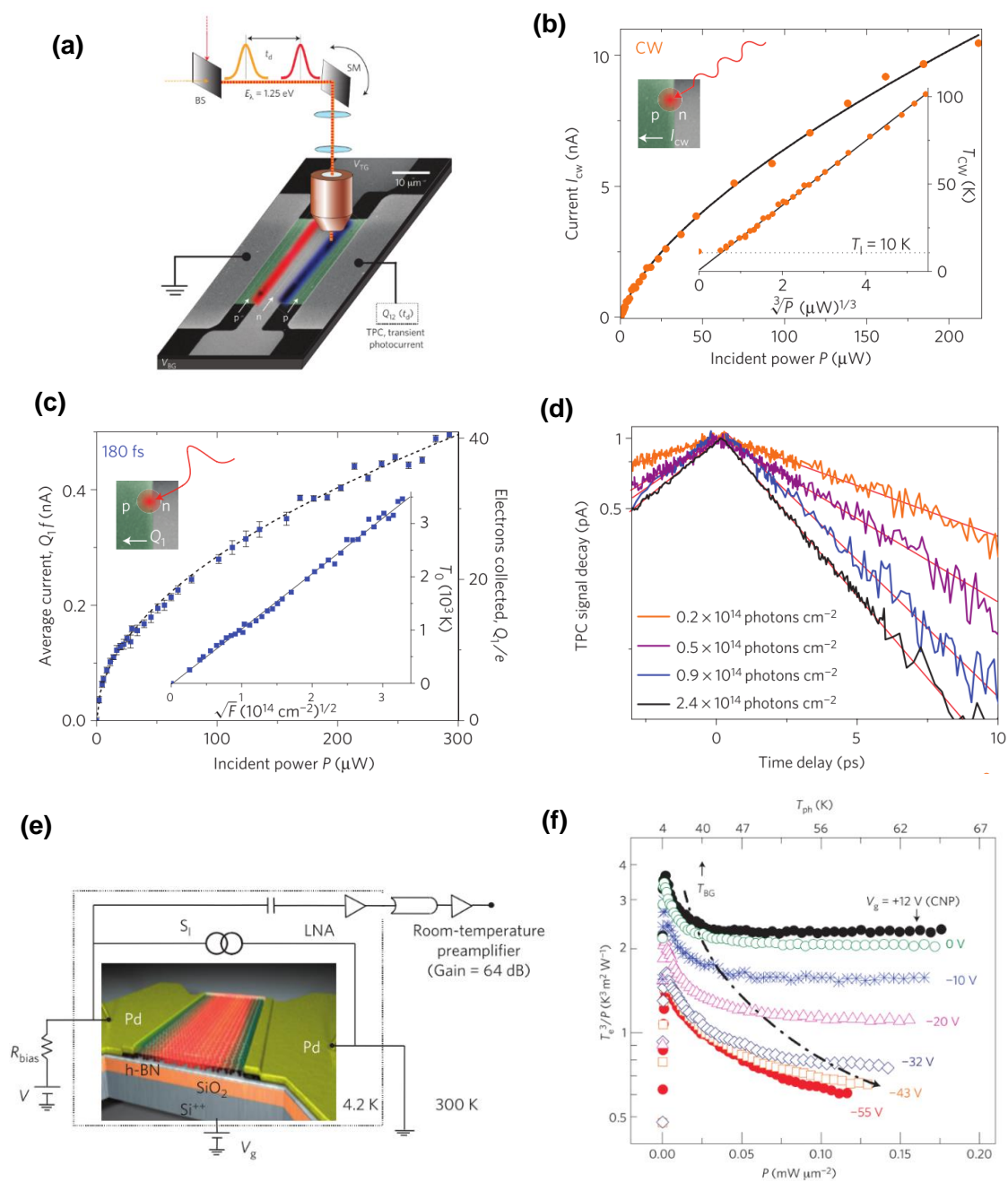

Figure 11. Experimental evidence of Supercollision cooling process. (a) Experimental configuration to measure the photocurrent as a function of laser power and time delay. (b) Photocurrent as a function of CW laser power at a graphene *p-n* junction (scattered points from experimental results, solid curve from fitting to the relation $I_{CW} \propto P^{0.65}$). (c) Averaged photocurrent as a function of pulsed laser power (scattered points) with a curve fitted to the relation $I_{PL} \propto P^{0.5}$. (d) Decay of transient photocurrent as a function of time delay for various incident laser power. (e) Experimental configuration for noise thermometry. (f) $T_{el}^3/P$ as a function of electrical power ($P$) for various gate voltages. $T_{BG}$ is the Bloch-Grüneisem temperature. Figures (a–d) are reproduced with permission from [109], © 2013 Macmillan Publisher Limited; (e),(f) from [112], © 2012 Macmillan Publisher Limited.

The direct measurement of $T_e$ in a graphene device to probe the electron-lattice cooling rate was also performed using Johnson noise thermometry, in which electrons are heated by a bias voltage [112]. The relationship between the Johnson noise and $T_e$ is given by $S_I = 4k_B T_e/R$, where $S_I$ is the current noise spectrum, $k_B$ is the Boltzmann constant and $R$ is the sample resistance. The experiments based on the noise thermometry have also successfully demonstrated that the energy relaxation rate is dominated by the supercollision process in the high temperature regime of $T_l \geq T_{BG}$ although it is governed by the standard electron-phonon scattering processes in the low temperature regime of $T_l < T_{BG}$ as shown in figures 11(e) and (f). Figure 11(f) shows the electron temperature, in the form of $T_e^3/P$, as a function of $P$. A $P \propto T_e^3$ behavior (following the supercollision cooling process) was observed at $T_l > T_{BG}$ and $P \propto T_e^4$ (following the conventional electron-AP scattering process) at $T_l < T_{BG}$. In figure 11(f), the dashed-dotted line indicates the gate-tunable $T_{BG}$. Here, $k_F l$ is estimated as ~3.5.

## 4.3 Cooling process in van der Waals structures

### *4.3.1 Out-of-plane heat transfer through electron-hyperbolic phonon coupling*

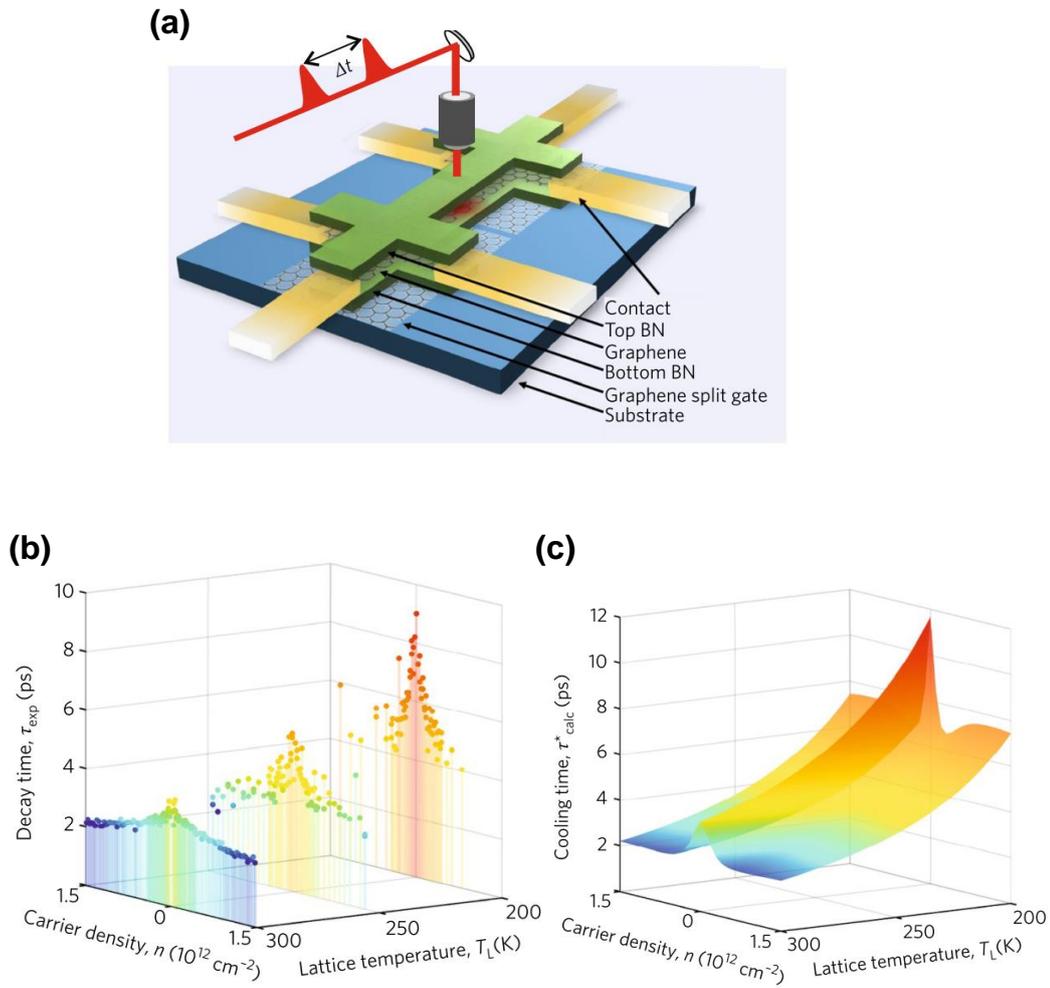

Figure 12. Electron-hyperbolic phonon coupling. (a) Schematic of transient photothermal voltage measurement for *h*-BN encapsulated graphene. (b) Experimental decay (cooling) time of electrons in graphene and (c) predicted near-equilibrium cooling time based on super-Planckian cooling to hyperbolic *h*-BN phonons. Figures are reproduced with permission from [95], © 2017 Macmillan Publisher Limited.

Cooling in *h*-BN-encapsulated graphene is governed by the out-of-plane coupling of graphene electrons to polar phonons in hBN (hyperbolic phonon polaritons). By varying the delay $t_d$ between two subpicosecond pulses, the carrier dynamics from the photovoltage signal $\Delta V_{PTE}(t_d)$ generated by the PTE effect were studied (figure 12(a)). The cooling timescale $\tau_{\exp}$ was estimated from the exponential decay, $\Delta V_{TPE}(t_d) \propto e^{-t_d/\tau_{\exp}}$. In this experiment, $k_F l$ was 80−100, corresponding to a mobility ($\mu$) of ~30,000 cm$^2$V$^{-1}$s$^{-1}$, compared to SiO$_2$-supported graphene devices in which $k_F l$ < 10 and $\mu$ < 5,000 cm$^2$V$^{-1}$s$^{-1}$. The exponential-decay behavior of the $\Delta V_{PTE}(t_d)$ with $\tau_{exp}$ ~ 2.5 ps at room temperature changes to a non-exponential one at low temperatures ( < 200 K). Figures 12(b) and (c) show that cooling becomes faster with increasing carrier density and $T_l$. For $h$-BN-encapsulated graphene, a heat transfer between hot graphene electrons and $h$-BN hyperbolic phonon polaritons becomes dominant in electron cooling. This is based on the absorption of thermal noise emitted from hot electrons by polar phonons in hBN. Due to the near-field coupling to hyperbolic modes, the hyperbolic heat transfer exceeds Planck's law by several orders of magnitude, a process known as super-Planckian coupling [113], where the cooling timescale with the energy transfer rate $Q$ is given by $\tau_{hc}(T_e, T_l) = C_e \frac{T_e - T_l}{Q}$. In the limit of weak heating (near-equilibrium state), an exponential decay with the time scale of $\tau_{hc}^* = C_e \left( \frac{\partial Q}{\partial T_e} |_{T_e \sim T_l} \right)^{-1}$ is obtained. It was shown that the hyperbolic hBN cooling model is consistent with experimental findings for carrier densities of $n < 1.6 \times 10^{12}$ cm$^{-2}$ and $T_l > 200$ K [95]. The model is however inapplicable at sufficiently low $T_l$ and higher $n$, where it can be assumed that graphene OPs and momentum-conserving cooling by scattering with APs dominate cooling. The electron-hyperbolic phonon cooling process was also observed using the noise thermometry method [114].

*4.3.2 Tuning ultrafast electron thermalization pathways in a van der Waals heterostructure*

The thermalization of photoexcited electrons through electron scattering occurs at extremely fast timescales of <30 fs, as mentioned above. Thus, the manipulation of the electron thermalization process is a challenging issue. In a vdW heterostructure consisting of

graphene/*h*-BN/graphene (G/*h*-BN/G) on a Si/SiO$_2$ substrate (see figure 13(a)), the photoexcited electrons in the top graphene layer can tunnel to the bottom graphene layer through the thin *h*-BN layer, resulting in interlayer charge transport before the onset of intralayer thermalization. In the experiment, the competing processes were observed by measuring the interlayer photocurrent ($I_{ph}$) under different bias and laser excitation conditions. $I_{ph}$ showed a linear or superlinear power dependence according to bias voltage ($V_{sd}$) and photon energy ($\hbar\omega$) conditions. At $V_{sd}$ = 5 V, $I_{ph}$ increases superlinearly with $P$ at $\hbar\omega$ = 1.75 eV (left panel), then becomes linear with increasing photon energy. In the right panel at $\hbar\omega$ = 2.1 eV, the superlinear curve changes to a linear one as $V_{sd}$ increases from 1.5 V to 10 V (figure 13(b)). With the results fitted according to the relation $I_{ph} \sim P^\gamma$, figure 13(c) displays $\gamma$ as a function of $V_{sd}$ and photon energy, showing a superlinear regime A ($\gamma > 1$) at low $V_{sd}$ and photon energy and a linear regime B ($\gamma \sim 1$). Figure 13(d) explains which process is dominant in each regime: thermionic emission in A and direct carrier tunneling in B. In regime A when the photoexcited carriers remain in the top graphene layer, they form a thermal equilibrium state, with a part of the carrier population having sufficient energy crosses the *h*-BN layer to the bottom graphene layer. The superlinear behavior comes from the exponential increase of the thermionically emitted carriers with temperature. In regime B, on the other hand, the excited carriers can directly tunnel to the other graphene layer through a reduced *h*-BN barrier with high $V_b$, based on the photon-assisted tunneling process. Here, $I_{ph}$ is proportional to the number of initial photoexcited carriers, resulting in a linear dependence on laser power. This indicates that the competition between the interlayer charge transport and intralayer thermalization can be tuned by the interlayer $V_{sd}$ or excitation photon energy.

The tunable photo-thermionic effect was also performed in a vertical G/WSe$_2$/G vdW heterostructures[115]. Thermalized hot carriers in the top graphene due to the absorption of photon energy flow to the bottom graphene through the WSe$_2$ by overcoming the Schottky barrier between graphene and WSe$_2$. Here, the number of carriers scales with a relation of $\exp(-\phi_B/k_B T_{el})$, where $\phi_B$ is the Schottky barrier height.

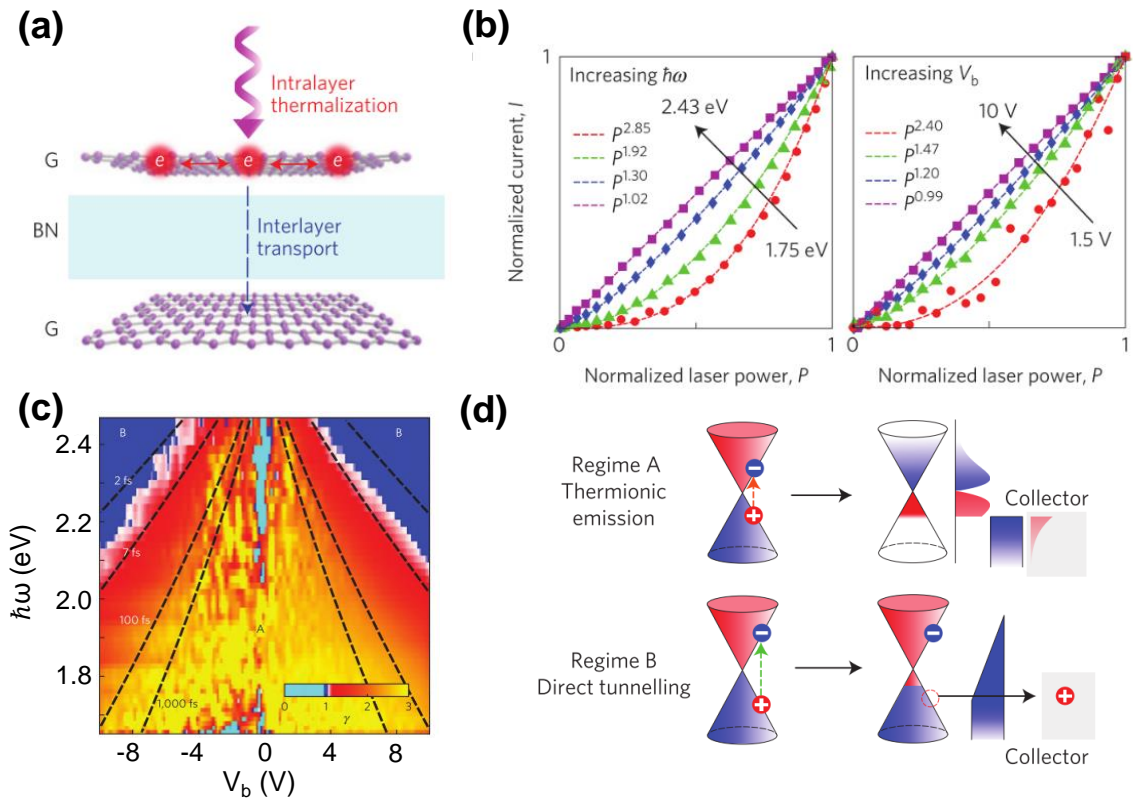

Figure 13. Tuning ultrafast electron thermalization pathways. (a) Schematic of interlayer thermalization and interlayer transport of optically exited electrons. (b−d) Two different regimes of interlayer photocurrent in a graphene/$h$-BN/graphene device. Figures are reproduced with permission from [116], © 2016 Macmillan Publisher Limited.

## 5. Summary and applications

While vdW heterostructures composed of graphene, $h$-BN, BP and various transition metal dichalcogenides have shown tremendous potential for electronic [117], optoelectronic [118] and spintronic [119] applications, understanding how hot electrons generated by high electric field and incident laser dissipate energy is important and can lead to the development of energy-efficient electronic and ultrafast optoelectronic devices based on physics unique to vdW heterostructures. Here, we discuss how the unique carrier scattering dynamics in graphene-based vdW heterostructures can be exploited for optoelectronic applications in light emission and photodetection.

**5.1 Energy dissipation in the steady state**

At steady state in vdW electronics, the energy dissipation of hot carriers plays a significant role in electrical properties such as current saturation [84] and depends on thermal properties such as thermal conductivity and thermal boundary resistance, including scattering with OPs in the substrate, which should be accounted for to fully explain the thermal and electrical behavior. In addition, hot electrons can dissipate energy through thermal radiation which potentially can be useful for applications like sensing, energy harvesting and lighting. Although thermal emission from an electrically biased graphene device has a rather broad spectrum following Planck's law, a photonic crystal nanocavity can be used to modify and control this thermal radiation spectrum. We can take advantage of the thermal decoupling of the hot electrons from the APs, which are much cooler, to use a Si-based photonic crystal nanocavity, which is unstable at temperatures above 1000 K, to modify the blackbody radiation from an electrically biased *h*-BN-encapsulated graphene emitter into and beyond the near-infrared spectrum. It was shown that such a device can emit visible light at $T_e \sim 2000$ K while the surrounding Si cavity remains at 700 K, the temperature of the APs [120].

**5.2 Energy dissipation in the dynamic state**

Photoexcited electrons in graphene undergo multiple cooling processes: electron-electron scattering, electron-OP scattering (including supercollision with impurities and APs), and electron-hole recombination. The supercollision process is found in both electrically biased graphene (steady state) and for optically excited carriers (dynamic state) and can be monitored in experiments through the resulting photocurrent. When photoexcited electrons are much hotter than the lattice, they can produce a photocurrent at a *p-n* junction based on the photo-thermoelectric effect [121]. It has been shown that the hot-carrier-dominant photocurrent response time is only 1.5 ps at room temperature corresponding to a bandwidth of ~500 GHz, which is limited by the charge-carrier cooling time [122]. At the *p-n* junction, the Peltier effect, as well as the OP-mediated cooling process, becomes an important cooling path. The Peltier cooling rate is proportional to the photo-thermoelectric effect. This indicates that the Peltier cooling as well as photocurrent due to the hot carriers at the junction can be manipulated by the junction configuration [122]. Also, the photovoltage generation time, related to the carrier heating event, was found to be faster than 50 fs, which implies a high carrier heating efficiency and that the *p-n* junction can be developed as an ultrafast femtosecond photodetector [123].

**Acknowledgments**

MHB was supported by the Korea Research Institute of Standards and Science (KRISS-GP2018-003), part of the Basic Science Research Program through the National Research Foundation of Korea (NRF) (Grant Nos. 2018R1A2A1A05078440 and SRC2016R1A5A1008184). He was also partly supported by the Korea-Hungary Joint Laboratory Program for Nanosciences through the National Research Council of Science and Technology. ZYO was supported in part by a grant from the Science and Engineering

Research Council (Grant No. 152-70-00017) and financial support from the Agency for Science, Technology and Research (A*STAR), Singapore.